\documentclass[aps,twocolumn,showpacs]{revtex4}
\usepackage{dcolumn}
\usepackage{graphicx}
\usepackage{amsmath}
\usepackage{amsfonts}
\usepackage{amssymb}
\usepackage{psfrag}
\usepackage{wrapfig}
\usepackage{subfigure}
\usepackage{makeidx}
\usepackage{bm}
\usepackage{epsf}

\begin{document}

\title{Vector rogue waves on a double-plane wave background}
\author{Li-Chen Zhao}
\author{Liang Duan}
\author{Peng Gao}
\author{Zhan-Ying Yang}\email{zyyang@nwu.edu.cn}

\address{$^{1}$School of Physics, Northwest University, Xi'an, 710069, China}
\address{$^{2}$Shaanxi Key Laboratory for Theoretical Physics Frontiers, Xi'an, 710069, China}

\date{\today}
\begin{abstract}
We  study rogue wave excitation dynamics on a double-plane wave background through deriving rogue wave solution on the background. The results indicate that rogue wave still can be excited successfully from resonant perturbations with the two plane wave backgrounds. The obtained vector rogue wave can be decomposed to two rogue waves located on the two backgrounds separately. This enables us to investigate the superpositions of two of the three well-known fundamental rogue wave patterns, mainly including eye-shaped, anti-eye-shaped, and four-petaled one. The explicit conditions for different possible superpositions are clarified by a phase diagram for rogue wave pattern on each plane wave background. The detail analysis indicate that the rogue wave admits many different profiles, in contrast to the ones reported before. The studies can be extended directly to investigate other localized waves on double-plane wave background and even more plane waves involved cases.

\end{abstract}
\maketitle

\emph{Introduction}--- Rogue wave (RW) depicts a
unique event that seems to appear from nowhere and disappear without
a trace, and can appear in a variety of different contexts
 \cite{Onorato,Kharif1,Osborne,Ruban,N.Akhmediev,Kharif,Pelinovsky}.  The
rational solutions on one plane wave background of related dynamics equations have been used to describe RW phenomena widely, which is supported by many experiments \cite{Solli, Kibler,Chabchoub,Bailung}. According to the component number, RW can be classified to scalar RW  \cite{R.Osborne,N.A,Yang,Ling,He,Ling2,Akhmediev}, and vector RW  \cite{Bludov,Baronio,Ling3,zhao2,Zhao3,Degasperis}. Most of previous RWs are excited on one plane wave background. But there are usually more than one plane wave backgrounds in complex systems. Then, if the background field contain more plane waves, can the RWs still exist or would the RW evolution characters be kept? We would like to study on this question through revisiting on the vector RWs, since the scalar RW system usually do not admit plane wave superposition background, and vector system could admit plane wave superposition form.

Vector RW has been studied widely on a plane wave background \cite{Baronio,Ling3,zhao2,Zhao3,Degasperis,Zhao4}. It has been shown that vector RWs admit more complex and abundant excitation patterns than the scalar ones. Anti-eye-shaped RW and four-petaled RW were reported in vector systems.  We still choose a standard two-component coupled nonlinear Schr\"{o}dinger equations (NLSE) to discuss the vector RW dynamics, because of its wide applications in physical systems \cite{Spinor,Agwal}. It is noted that the coupled NLSE can admit cosine or sine type seed solution, which can be decomposed to a  double-plane wave background (DPWB). This provides possibilities to investigate RWs on a DPWB in detail.

In this paper,  we obtain the vector RW excitations on a DPWB with the aid of Darboux transformation. The results indicate that RW still can be excited successfully from resonant perturbations with the two plane wave backgrounds. The obtained vector RWs can be decomposed to two RWs located on the two backgrounds separately. This enables us to investigate the superpositions of two of the three well-known fundamental RW patterns, mainly including eye-shaped, anti-eye-shaped, and four-petaled one. The detail analysis indicate that the RWs on two plane wave background are related, and there is at least one eye-shaped RW on one of the two plane waves. It is impossible to obtain two anti-eye-shaped RWs  or two four-petaled RWs  on the DPWB. The possible combination of them are clarified clearly by a phase diagram. Moreover, further analysis indicate that there are two fundamental RW structures on the same background, in contrast to the only one fundamental eye-shaped structure on a background. The two fundamental RW structures come from the two branches of modulational instability dispersion form on a certain background.

 \emph{The coupled nonlinear Schr\"{o}dinger equations and double-plane wave backgrounds}---
 We begin with the well known two-coupled NLSE in dimensionless form
\begin{equation}\label{cnls}
    \begin{split}
      i q_{1,t}+\frac{1}{2}q_{1,xx}+\frac{1}{2}(|q_1|^2+|q_2|^2)q_1 &=0,  \\
      i q_{2,t}+\frac{1}{2}q_{2,xx}+\frac{1}{2}(|q_1|^2+|q_2|^2)q_2 &=0.
    \end{split}
\end{equation}
The coupled NLSE model can describe the dynamics of matter wave in quasi-1 dimensional two-component Bose-Einstein condensate \cite{Spinor}, the evolution of optical fields in a two-mode or polarized nonlinear fiber \cite{Agwal}, and even the vector financial system \cite{Yan}. The vector RW have been studied widely on plane wave backgrounds, for which there is only one plane wave background in each component \cite{Baronio,Ling3,zhao2,Zhao3,Degasperis}. The vector RW demonstrates many nontrivial properties compared with the scalar RW of NLSE. They admit two more fundamental RW patterns, mainly including anti-eye-shaped, and four-petaled one, which are quite different from the eye-shaped one obtained widely in scalar RW. We would like to check if the RW can exist on a DPWB through deriving the rational solution on the DPWB as previous works \cite{R.Osborne,N.A,Yang,Ling,He,Ling2,Akhmediev,Baronio,Ling3,zhao2,Zhao3,Degasperis}.

It is found that the CNLSE also admit the following DPWB solution except the well known  one plane wave background, the DPWB can be written as
\begin{equation}\label{cnls}
    \begin{split}
      q_{10}&=a_1 e^{{\rm i}(k_1 x-k_1^2 t/2+\phi\ t)}+a_2 e^{{\rm i} (k_2 x-k_2^2t/2+\phi \ t)},  \\
     q_{20}&=a_1 e^{{\rm i}(k_1 x-k_1^2t/2+\phi \ t)}-a_2 e^{{\rm i} (k_2 x-k_2^2t/2+\phi \ t)},
    \end{split}
\end{equation}
where $\phi=a_1^2+a_2^2$ is the nonlinear effects induced phase evolution factor.  $a_i$ and $k_i$ are the amplitude and wave vector of background respectively. When $k_1=k_2$, the DPWB will be reduced to be one plane wave background; when $k_1=-k_2=k$, the DPWB will be reduced to be a cosine or sine wave background. The periodic backgrounds are similar to the stripe cases studied in a two component BEC with pair-transition effects \cite{Zhao5}. As far as we know, the stripe phase background is  obtained for the first time in a two-component BEC with no transition effects. If $|k_1|\neq |k_2|$, the background usually describe a moving stripe background, namely, the background density also depend on time. These characters enable us to investigate vector rogue waves on a plane wave background as previous studies, and vector rogue waves on a DPWB.  Since the DPWB reduces to be one plane wave background for $k_1=k_2$ cases, we mainly discuss the RW excitation patterns on DPWB with different wave vectors. Firstly, let us discuss on the fundamental vector RW on the DPWB, which is the first step to describe and understand the RWs on DPWB.

\emph{Fundamental vector rogue wave on a double-plane wave background}---

The fundamental RW solution on the DPWB for above vector NLSE can be presented as
\begin{widetext}
\begin{equation}\label{fundrw}
\begin{split}
q_{1F}&=q_{10}+\frac{2{\rm i}[a_1 A_2 {\rm e}^{\theta_1}(\chi_{R}+k_1)+a_2 A_1 {\rm e}^{\theta_2}(\chi_{R}+k_2)](x+\chi_{R}t)-(a_1 A_2 {\rm e}^{\theta_1}+a_2 A_1 {\rm e}^{\theta_2} )(2{\rm i}\chi_{I}^2t+1)}{A_1 A_2 [(x+\chi_{R}t)^2+\chi_{I}^2t^2+\frac{1}{4\chi_{I}^2}]},  \\
     q_{2F}&=q_{20}+\frac{2{\rm i}[a_1 A_2 {\rm e}^{\theta_1}(\chi_{R}+k_1)-a_2 A_1 {\rm e}^{\theta_2}(\chi_{R}+k_2)](x+\chi_{R}t)-(a_1 A_2 {\rm e}^{\theta_1}-a_2 A_1 {\rm e}^{\theta_2} )(2{\rm i}\chi_{I}^2t+1)}{A_1 A_2 [(x+\chi_{R}t)^2+\chi_{I}^2t^2+\frac{1}{4\chi_{I}^2}]},
\end{split}
\end{equation}
\end{widetext}

where $A_i=(\chi_{R}+k_i)^2+\chi_{I}^2$, $\theta_i={\rm i}\left [k_i x+(a_1^2+a_2^2-k_i^2/2)t\right]$, $i=1,2$. The parameter $\chi_{R}=\mathrm{Re}(\chi)$, $\chi_{I}=\mathrm{Im}(\chi)$, and  $\chi$ is a root for the following equation
$1+\sum_{i=1}^{2}\frac{a_i^2}{(\chi+k_i)^2}=0$. The solution enables us to investigate RW dynamics on a DPWB. For an example, we show on case in Fig. 1. It is shown that the RW indeed exist on the DPWB, and the backgrounds demonstrate both spatial and temporal oscillation behaviors. The RW structure is distinctive from the ones reported before \cite{Baronio,Ling3,zhao2,Zhao3,Degasperis}.  The superposition RWs admit about $1.6$ times background amplitude in both the two components. Fundamental RW for scalar NLSE system always admit peak value three times background amplitude. The RW's peak value is much smaller than the RW on one plane wave background with an identical background density value reported before. The RWs peak is found to depend on relative wave vector $k_2-k_1$ and the ratio of two background amplitudes $a_2/a_1$.  Nextly, let's discuss which types of RW dynamics can be investigated based on the obtained RW solution.
\begin{figure}[htb] \centering
\includegraphics[height=40mm,width=85mm]{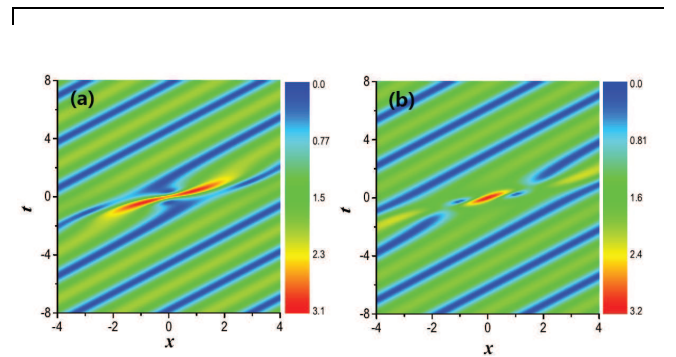}
\caption{(color online) The rogue wave excitations on a double-plane wave background. (a) for rogue wave pattern in component $q_1$ and (b) for rogue wave in component $q_2$.  Based on the phase diagram in Fig. 2, we can know that the rogue wave in each component is the results of linear superposition of an eye-shaped and anti-eye-shaped one on the two plane waves separately.  The parameters are $a_1=a_2=1$, $k_1=0$, $k_2=2$.}
\end{figure}

\begin{figure}[htb] \centering
\includegraphics[height=40mm,width=85mm]{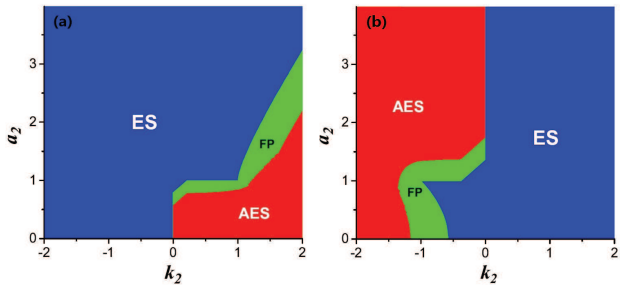}
\caption{(color online) A phase diagram for rogue wave pattern on a double-plane wave background. (a) for rogue wave structure types on one plane wave background $a_1 e^{{\rm i}(k_1 x-k_1^2 t+\phi\ t)} $ and (b) for rogue wave structure types on the other plane wave background $ a_2 e^{{\rm i} (k_2 x-k_2^2t+\phi \ t)}$. ``ES", ``AES", and ``FP" denote eye-shaped, anti-eye-shaped, and four-petaled structure for rogue wave pattern respectively. This phase diagram enables one to know which conditions for superpositions of two of the three fundamental structures. The parameters are $a_1= 1$, $k_1=0$.}
\end{figure}

 The obtained vector RW can be decomposed to two RWs located on the two plane wave backgrounds separately, explicitly, $q_{1F}=H_1(x,t)+H_2(x,t)$, $q_{2F}=H_1(x,t)-H_2(x,t)$, where $H_i=a_i{\rm e}^{\theta_i}+a_i\frac{2{\rm i}(\chi_{R}+k_i)(x+\chi_{R}t)-2{\rm i}\chi_{I}^2t-1}{A_i [(x+\chi_{R}t)^2+\chi_{I}^2t^2+\frac{1}{4\chi_{I}^2}]}{\rm e}^{\theta_i}$.  The RW on each plane wave background has been discussed clearly. They mainly admit three well-known fundamental RW patterns, mainly including eye-shaped, anti-eye-shaped, and four-petaled one. This enables us to investigate the linear superposition of two of the three well-known fundamental RW patterns. It has been shown that the RW structure depends on the background field properties. The anti-eye-shaped RW and MI character in vector NLSE have been observed in a two-mode nonlinear fiber \cite{Frisquet,Frisquet2}. Under which conditions for observing the superposition of two fundamental patterns should be discussed in detail for systemic experimental observation.

 Recently, baseband MI or MI with
resonant perturbations was found to play an essential role in RW excitations
\cite{Baronio1,Baronio3,zhaoling}. Through Fourier analysis, we can see that each perturbation is resonant with the plane wave background on which it exist. This point agrees well with that the resonant perturbation in MI regime induce RW excitation \cite{zhaoling}. Furthermore, the underlying mechanisms for forming different spatial-temporal structures of fundamental RWs  have been uncovered \cite{Lingzhao}.
 Since the $\chi$ is a root of a fourth-order equation, there are two MI branches which is different from the one MI branch for scalar NLSE. This makes there are two differen patterns on a identical background, which provides possibilities to investigate nonlinear superposition of different RW patterns on one plane wave background  \cite{Lingzhao3,Chen}. The two fundamental RW structures come from the two branches of MI dispersion form on a certain background. Therefore, we plot the conditions for different fundamental patterns according to one MI branch. The other MI branch can be addressed similarly. Through analyzing the extreme points of the RW solution on spatial-temporal distribution plane \cite{Lingzhao}, we summarize the conditions for these different fundamental patterns according to one branch (shown in Fig. 2). It should be noted that the perturbation signals for RWs on the two plane wave backgrounds are connected by the condition $ 1+\sum_{i=1}^{2}\frac{a_i^2}{(\chi+k_i)^2}=0 $. Therefore, we plot the RW patterns on the plane wave backgrounds separately (Fig. 2(a) and (b)). The phase diagram is plot in a two-parameter space which are relative amplitude $c_2$ and wave vector $k_2$ with $c_1=1$ and $k_1=0$. The settings of $c_1=1$ and $k_1=0$ do not losing generality, because of the scalar transformation invariable and relative wave vector properties.  It is seen that eye-shaped one always exist at least in one component. The regime for four-petaled RW is much smaller than the ones for eye-shaped one and anti-eye-shaped one. This explains well why the four-petaled RW was not shown in the two-component case \cite{zhao2}, and it was demonstrated by varying relative frequency between the two components \cite{Zhao4}. From the phase diagram, we can see that it is possible to obtain superposition of eye-shaped one with eye-shaped, anti-eye-shaped or four-petaled one. It is impossible to obtain superposition of anti-eye-shaped one with anti-eye-shaped one or four-petaled one in the this two component case. Meanwhile, we emphasize that the results are discussed for the integrable focusing-focusing NLSE case.  The similar discussions should be made in detail for other cases, such as un-integrable case, defocusing-defocusing, and mixed cases.

 \begin{figure}[htb] \centering
\includegraphics[height=40mm,width=85mm]{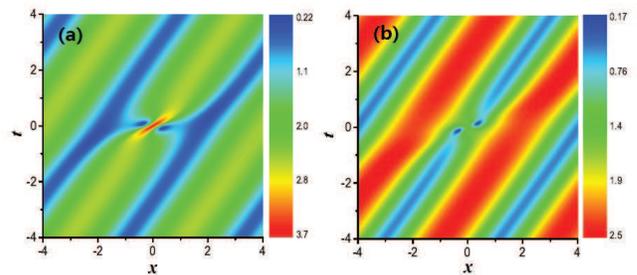}
\caption{(color online) The rogue wave excitations on a double-plane wave background. (a) for rogue wave pattern in component $q_1$ and (b) for rogue wave in component $q_2$.  Based on the phase diagram in Fig. 2, we can know that the rogue wave in each component is the results of linear superposition of an eye-shaped and a four-petaled one on the two plane waves separately.  The parameters are $a_1=1, a_2=1.5$, $k_1=0$, $k_2=1.5$.}
\end{figure}

 \begin{figure}[htb] \centering
\includegraphics[height=36mm,width=85mm]{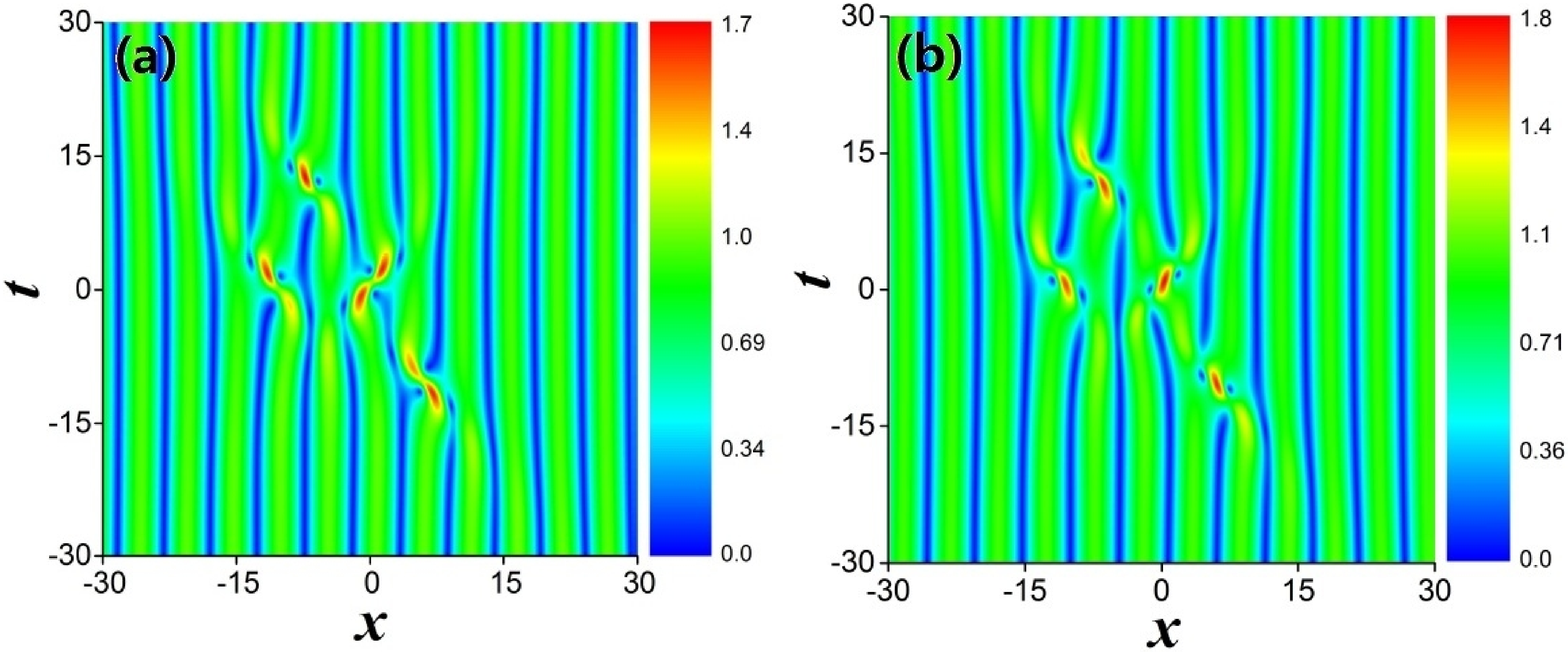}
\caption{(color online) One case of four rogue wave excitations on a double-plane wave background. (a) and (b)  for rogue waves in the component $q_1$ and component $q_2$ respectively.    It is shown multi-rogue wave can be also exist on the double-plane wave background. The parameters for backgrounds are $a_1= a_2=1/2$, $k_1=3/5$, $k_2=-3/5$.}
\end{figure}

 \begin{figure}[htb] \centering
\includegraphics[height=36mm,width=85mm]{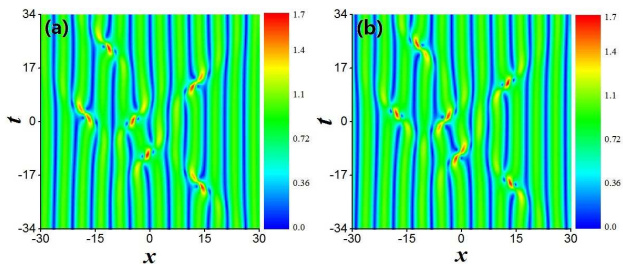}
\caption{(color online) One case of six rogue wave excitations on a double-plane wave background. (a) and (b)  for   rogue waves in the component $q_1$ and component $q_2$ respectively.    It is shown multi-rogue wave can be also exist on the double-plane wave background. The parameters for backgrounds are $a_1= a_2=1/2$, $k_1=3/5$, $k_2=-3/5$.}
\end{figure}

These results enable us to clarify the explicit conditions for investigating different superposition of fundamental RWs. For example, Fig. 1 is plotted with $a_1=a_2=1$, $k_1=0$, $k_2=2$ according the identical branch for Fig. 2. Based on the phase diagram, we can know directly that the RW in Fig. 1 is the results of linear superposition of an eye-shaped and anti-eye-shaped one.  If we want to investigate the superposition of an eye-shaped and a four-petaled one, we can choose the background condition $c_2= 1.5 $ and $k_2= 1.5 $. The dynamical processes in the two components are shown in Fig. 3.

After the Peregrine RW was observed experimentally in many
different physical systems, high-order RW were excited successfully
in a water wave tank, mainly including the RW triplets for second
-order one \cite{Chabchoub2} and up to the fifth-order ones
\cite{Chabchoub3}. This suggests that high-order analytic RW
solution is meaningful physically and can be realized experimentally
\cite{Erkintalo}. Therefore, we present some cases for dynamics of high-order vector RWs on the DPWB.

\emph{Multi-vector rogue wave on a double-plane wave background}----
The high-order vector RWs have been derived in \cite{Lingzhao2,Chen,Lingzhao3}. The vector RWs including multi-ones and high-order ones have been derived and investigated in details \cite{Lingzhao2}.  Similarly, we can derive the muti-RWs and high-order ones on the DPWB by performing the methods \cite{Lingzhao2}.  Because the related expressions for them are too complicated, we do not show them here. For examples, we show four RWs on the background in Fig. 4, and six RWs on the DPWB in Fig. 5. These results indicate that the perturbations with high-order resonance in modulational instability can excite high-order RWs on the DPWB. It is emphasized that the four RWs  distribution profiles on temporal-spatial plane are distinctive from the four RWs  reported in  \cite{Lingzhao2}, since the four RWs is a superposition of fundamental RW and a second-order RW. The six RWs are a superposition of two second-order RWs, it is different from the six RWs for third-order RW in scalar NLSE systems \cite{Ling2,Akhmediev}.

\emph{Conclusions}---  We obtain the vector RW excitations on a general DPWB with the aid of Darboux transformation. The results indicate that RW still can be excited successfully from resonant perturbations with the two plane wave backgrounds. The obtained vector rogue wave can be decomposed to two RWs located on the two backgrounds separately. This enables us to investigate the superposition of two of the three well-known fundamental RW patterns. The explicit conditions for different possible superpositions are clarified by a phase diagram for rogue wave pattern on each plane wave background. The studies would inspire more studies to investigate other localized waves on double-plane wave background and even more plane waves involved cases, and provide possibilities to generate RW on multi-plane wave background in experiments. 

Very recently, it was shown that the integrable two-component coupled NLSE  admitted $SU(2)$ symmetry and proper linear combinations of the solutions of  the integrable coupled NLSE were also the solutions of the coupled NLSE \cite{Kartashov}. The solutions here can be also obtained by the linear transformation. Meanwhile, it is emphasized that the multi-component coupled NLSE admit similar symmetry, and can be used to construct many other different nonlinear waves on multi-plane wave background. These properties will be used to direct experimental observations on RW on multi-plane wave backgrounds, which are more practical in real complex systems.

\section*{Acknowledgments}
This work is supported by National Natural Science Foundation of
China (Contact No. 11775176,11405129), and Shaanxi Province Science association of colleges and universities (Contact No. 20160216).

\end{document}